# Code Quality Evaluation methodology using the ISO/IEC 9126 Standard

Yiannis Kanellopoulos[1], Panos Antonellis[2], Dimitris Antoniou[2], Christos Makris[2], Evangelos Theodoridis[2], Christos Tjortjis*[3,4], and Nikos Tsirakis[2]

[1] University of Manchester, School of Computer Science, Manchester, U.K.
ykanell@gmail.com
[2] Dept. of Computer Engineering and Informatics, University Of Patras, Greece
{adonel,antonid,makri,theodori,tsirakis}@ceid.upatras.gr
[3] Dept. of Computer Science, Univ. of Ioannina, P.O. 1186, 45110, Ioannina, Greece
[4] Dept. of Engineering Informatics and Telecoms, University of W. Macedonia, Greece
Christos.Tjortjis@manchester.ac.uk

## ABSTRACT

*This work proposes a methodology for source code quality and static behaviour evaluation of a software system, based on the standard ISO/IEC-9126. It uses elements automatically derived from source code enhanced with expert knowledge in the form of quality characteristic rankings, allowing software engineers to assign weights to source code attributes. It is flexible in terms of the set of metrics and source code attributes employed, even in terms of the ISO/IEC-9126 characteristics to be assessed. We applied the methodology to two case studies, involving five open source and one proprietary system. Results demonstrated that the methodology can capture software quality trends and express expert perceptions concerning system quality in a quantitative and systematic manner.*

## KEYWORDS

*Software Quality Management, Static Analysis, Software Metrics, ISO/IEC 9126*

## 1. INTRODUCTION

Software systems are large, complex beset with maintenance problems, whilst users expect high quality [1]. However, it is hard to assess and assure quality. The ISO/IEC 9126 standard has been developed to address software quality issues [2]. It specifies software product quality characteristics and sub-characteristics and proposes metrics for their evaluation. It is generic, and can be applied to any type of software product by being tailored to a specific purpose [3].

This work focuses on source code internal quality evaluation. Its contribution is a methodology for the software product quality assessment, using the ISO/IEC-9126 standard as a frame of reference and a set of metrics extracted solely from source code. The methodology comprises a three-step approach and a model that links system level quality characteristics to code-level metrics, and the application of the Analytic Hierarchy Process (AHP) [4] to every level of the model's hierarchy, in order to reflect the importance of metrics and system properties on evaluating quality characteristics.

Two case studies were conducted in order to evaluate the methodology. Open source and proprietary systems of different functionality, volume and development paradigm were used. Experimental results showed that it is able to detect software quality characteristic evolution and express expert perceptions concerning system quality and maintainability in a quantitative and systematic manner.

The remaining of this paper is organized as follows: Section 2 reviews software quality models and motivates the selection of the ISO/IEC 9126 standard for this work. Section 3 outlines our method for defining source code attributes, their metrics and weights reflecting their importance when evaluating ISO/IEC-9126 characteristics. Section 4 reviews results from application of the proposed methodology and assesses its accuracy. Finally, Section 5 concludes the paper.

## 2. BACKGROUND

Quality is the "totality of characteristics of an entity that bear on its ability to satisfy stated and implied needs" [2]. By employing the term "satisfaction", ISO/IEC 9126 implies "the capability of the software to satisfy users in a specified context of use". ISO/IEC 9126 Software Engineering – Product Quality Standard assesses a system's internal and external quality, as well as quality in use.

We base our methodology on internal quality, which consists of six characteristics: a) *functionality*, concerned with what the software does to fulfil user needs, b) *reliability*, evaluating software's capability to maintain a specified level of performance, c) *usability*, assessing how understandable and usable the software is, d) *efficiency*, evaluating the capability of the software to exhibit the required performance with regards to the amount of resources needed, e) *maintainability*, concerned with the software's capability to be modified, and f) *portability*, measuring the software's capability to be transferred across environments.

### 2.1. Software Quality Models

Several models which employ a set of quality attributes, characteristics and metrics were developed for this purpose. Chidamber and Kemerer presented the Metrics for Object-Oriented (OO) Software Engineering (MOOSE) suite [5]. It consists of six metrics that assess different OO system attributes: Weighted Methods per Class (WMC) measures complexity, Lack Of Cohesion in Methods (LCOM) measures cohesion, Coupling Between Objects (CBO) measures coupling, Depth of Inheritance Tree (DIT) and Number Of Children (NOC) assess inheritance, and finally Response For a Class (RFC) assesses messaging. This empirically validated work pioneered OO system quality evaluation, but offers no clear and explicit relationship between metrics and software system quality characteristics [6].

NASA's Software Technology Assurance Center (SATC), has developed the SATC Software Quality Model for Risk Assessment [7]. This model includes goals, associated attributes and metrics. It supports risk management and quality assessment of software process and products. The Quality Model for OO Design (QMOOD) is a comprehensive quality model that establishes a clearly defined and empirically validated model which assesses OO quality characteristics, such as understandability and reusability, and relates these to structural OO design attributes, such as encapsulation and polymorphism [8]. It also provides metrics and a method to quantify these attributes. The QMOOD model consists of six equations that establish relationships between six quality characteristics (reusability, flexibility, understandability, functionality, extendibility, and effectiveness) and eleven design properties.

The models presented above have characteristics similar to ISO/IEC-9126, in the sense that they indicate relationships between metrics and system quality characteristics. However, ISO/IEC-9126, unlike these models, enjoys the benefit of being an international standard agreed upon by the community. It defines a common language relating to software product quality and is widely recognized as such [9]. Thus we selected ISO/IEC-9126 as a frame of reference for our software product quality assessment methodology.

Nevertheless, ISO/IEC 9126 is not flawless: it has been criticized as difficult to be made operational and non-practical. It suggests metrics that are not based on direct observation of the software product, but rather of the interaction between the product and its environment (maintainers, testers and administrators) or on comparison of the product with its specification, which itself could be incomplete, out of date, or incorrect [10]. Moreover, the metrics provide guidance for *a posteriori* evaluation of the quality characteristics based on effort and time spent on activities related to the software product, such as maintenance [11]. Unfortunately, such metrics do not provide predictive power for quality characteristics. Neither does ISO/IEC-9126 provide guidance on how to weigh or collate metrics criteria in order to reflect their importance concerning quality factors.

In order to address those issues several refinements of ISO/IEC-9126 have been proposed. The "Evaluation Method for Internal Software Quality" (E.M.I.S.Q.) was recently presented [12]. It consists of a quality and assessment model and a tool. The proposed model consists of five quality attributes and three to nine sub-attributes per attribute. It assesses quality attributes and sub-attributes, and overall quality. On the level of sub-attribute assessment, an expert evaluator rates on a three-valued scale (ok, critical, very critical) how the results of the static code analysis tools are affecting them. On the level of quality attributes, the expert checks the plausibility of the available ratings as calculated median values.

Another model is proposed in [11], which maps a set of source-code measures onto ISO/IEC-9126 maintainability sub-characteristics, following pragmatic mapping and ranking guidelines. More specifically, the proposed model links system-level maintainability characteristics to code-level measures in two steps: firstly, it maps these system-level characteristics to source code level properties, (e.g. system changeability is mapped to source code complexity), and secondly, one or more source code measures are determined for each property, (e.g. source code complexity is measured in terms of cyclomatic complexity). The common features of these approaches are that: a) they attempt to map metrics onto ISO/IEC-9126's sub-characteristics and characteristics in a non quantitative way following the standard's proposed hierarchy; b) they lack a sophisticated way for eliciting weights for metrics and sub-characteristics in order to reflect their importance when evaluating ISO/IEC-9126 characteristics. However, there is no consensus among these two approaches on how to select quality metrics in order to assess ISO/IEC-9126 sub-characteristics and characteristics.

## 2.2. Contribution

We propose a source code quality evaluation methodology based on the ISO/IEC-9126 software quality standard. Similarly to [11] it enhances the hierarchy of the model by introducing a layer which reflects a software system's source code attributes, such as volume, size, complexity, cohesion and coupling. These attributes directly influence ISO/IEC-9126 characteristics and sub-characteristics. The difference is that our work does not focus only on maintainability, but also on portability, functionality and efficiency. In this way it is similar to [12], which however does not use the intermediate layer of source code properties.

We employ AHP for the assignment of relative weights to the employed metrics and source code attributes in order to reflect their importance on evaluating ISO/IEC-9126's characteristics and sub-characteristics. By using this multicriteria decision making method, and certain subjective elements, such as expert opinion, the proposed methodology makes accurate recommendations that approximate the model's entity relationships.

## 3. PROPOSED METHODOLOGY

As the scope of our methodology is to evaluate code quality with metrics calculated using input elements extracted solely from source code, some of the six ISO/IEC-9126 characteristics do not fit our purposes. More specifically, reliability, usability and *time behaviour*, a sub-characteristic of efficiency, are excluded, as they are more related to dynamic system behaviour. Therefore, internal quality is assessed based on four characteristics (functionality, efficiency, maintainability, portability) and their respective sub-characteristics. These are evaluated by employing a set of metrics. For instance, the quality level for maintainability takes into account the measured values of four sub-characteristics. By aggregating the values of these sub-characteristics a single value on a composite measure of maintainability can be derived. The above quality characteristics are abstract concepts and therefore not directly measurable and observable. Each of them is characterized by a set of sub-characteristics that are presented in Fig. 1. This final set of quality characteristics is generic enough to satisfy the main goal of this work, which is to create a model that supports system quality evaluation.

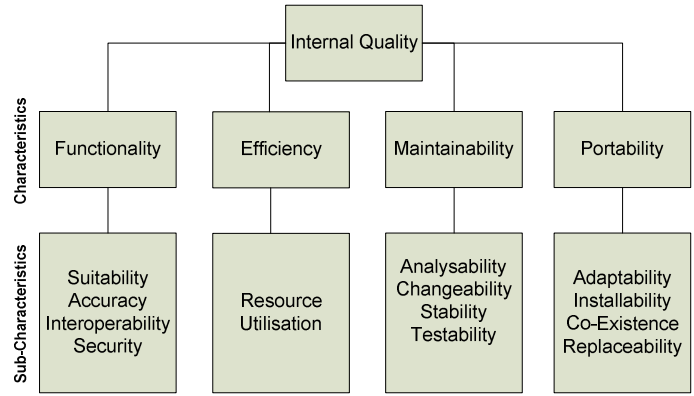

Figure 1. ISO/IEC 9126 Internal Quality Characteristics and Sub-Characteristics

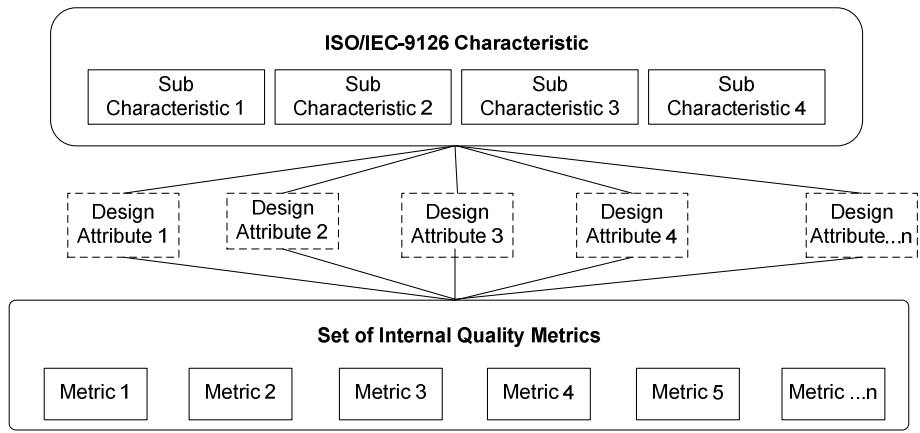

Figure 2. Hierarchy for the Proposed Methodology

## 3.1. Hierarchy of Characteristics, Attributes and Metrics

The proposed methodology is flexible in order to facilitate addressing varying perspectives, goals and objectives. Fig. 2 depicts the proposed hierarchy, where at the lowest level are metrics for assessing source code attributes, assigned to ISO/IEC-9126 characteristics and sub-characteristics. The methodology enables employing different sets of metrics or source code attributes, even changing the subset of ISO/IEC-9126 characteristics to be assessed.

### 3.1.1. Source Code Attributes

Source code attributes directly influence ISO/IEC 9126 characteristics and sub-characteristics [2]. They are tangible concepts that can be directly assessed by examining the static behaviour, relationship and functionality of software artefacts. For example, an evaluation of a class's coupling and the examination of its member methods and data reveals significant information concerning its structural and behavioural characteristics.

The OO approach involves source code attributes similar to the structured approach [8]. These attributes are *abstraction*, *encapsulation*, *coupling*, *cohesion, complexity* and *volume*. However, attributes such as *messaging*, *composition*, *inheritance* and *polymorphism* represent concepts that have been introduced by the OO paradigm and are vital for the OO system quality assessment [8], [4]. Table 1 presents the source code attributes that influence ISO/IEC 9126 characteristics, along with their definitions.

Table 1. Source Code Attributes

| Attribute | Definition |
|---|---|
| Volume | It measures the volume of the system [13]. |
| Complexity | It is determined in terms of structural characteristics by examining how objects are interrelated [13]. |
| Abstraction | The ability to engage with a concept, whilst safely ignoring some details [13]. |
| Encapsulation | The grouping of related concepts into one element [13]. |
| Coupling | The connections between objects (how tightly objects are related) [13]. |
| Cohesion | The degree to which an object has all operations working together in order to achieve a single, well-defined purpose [14]. |
| Messaging | Collaboration between objects by message exchange [13]. |
| Polymorphism | The ability of different objects to respond to the same message in different ways, enabling objects to interact without knowing their exact type [15]. |
| Composition | A strong form of aggregation where the "whole" is completely responsible for its parts and each "part" object is only associated to one "whole" object [2]. |
| Inheritance | A measure of relationships such as "is a" and "is like" [13]. |

### 3.1.2. Selected Metrics

Software demonstrates regular behaviour and trends, which can be measured [16]. Software quality assessment requires the collection of such metrics in order to provide a systematic approach of code evaluation based on a set of predefined rules. These metrics can also be useful as indicators for identifying potential problematic areas. Each of the source code attributes, mentioned in the previous section, can be objectively assessed using one or more well-defined metrics.

### 3.2. Software System Entities

A key feature of the proposed methodology is a model which models source code attributes and metrics in order to assess static internal quality. The definition of this model requires specification of software system entities and their respective attributes. Entities were selected to be applicable to all types of software systems, and easily identifiable, i.e. clearly named in a program. For instance, methods or functions are present in any type of software system, whereas classes exist only in OO systems. These entities should also model a large proportion of the code, thus ensuring that any subsequent analysis covers a large part of a system under evaluation. Moreover, entities need to contain a common set of attributes in order to enable uniform assessment. This allows for entity evaluation on the basis of their attributes. The number of selected attributes needs also to be sufficient in order to avoid misleading comparisons. Selected attributes can be both quantitative and qualitative [17], [18].

Software systems employ larger units of abstraction in order to separate the development of different constituents. Concepts such as the C *code-file*, Java *package*, C++ *namespace* and Delphi *unit* facilitate the development of isolated modules and cater for programmers to write their own applications in an abstract way [20]. In the case of systems developed using the structured approach the most important structures are functions (or subroutines, methods, procedures and blocks) [19]. On the other hand the design of an OO system is defined by modules (packages in Java, namespaces in C++ and units in Delphi), classes, and the relationships between them.

A function (or member method in OO systems) constitutes of input parameters, variables and its body of statements. A set of attributes that play an important role in the evaluation of the quality of a function includes: name, parameter types, number of parameters, parameters passing (by value, by reference), return type, complexity, coupling, number of statements, number of comments, and size. In case of an OO system methods are also evaluated based on their visibility; that is if they are public, protected or private. Functions are processing data variables in order to perform the required operations. Therefore, data variables are the most fundamental components in a software system. A set of attributes that influence the evaluation of the data variables is: name, type and kind (primitive or user defined).

A class constitutes of member data which represent its attributes. A set of operations (methods) that operates on a class' member data constitute its member methods. Another important characteristic of OO systems are the class hierarchies that organize groups of related classes. Member data form the basis of an OO system; the set of attributes that can influence the evaluation of their quality includes: name, visibility (public, protected, private), qualifiers (static, constant) and type.

Classes are evaluated based on their constituents (i.e. member methods and data) and their interactions with other classes. A set of attributes that can influence the evaluation of classes' quality includes: name, visibility (public, protected, private), name of ancestor, package/namespace/unit name, number of children, number of methods (public, protected, private), depth of inheritance tree, coupling, cohesion, complexity, number of statements and number of comments. A set of attributes that can influence the evaluation of classes' quality includes: name, afferent (inward) and efferent (outward) couplings. Fig. 3 presents the hierarchy of entities of a software systems' quality evaluation model.

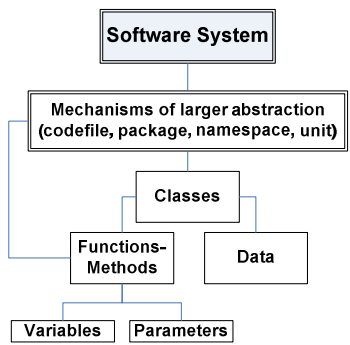

Figure 3. Entity Hierarchy

### 3.3. Employing AHP for weights assignment

The proposed methodology employs AHP at every level of its hierarchy, as shown in fig. 2. At the lower level we evaluate source code attributes using metrics (volume, complexity, etc.). We apply AHP also at the intermediate level to evaluate sub-characteristics using source code attributes. Finally, quality characteristics are evaluated at the top level from their respective sub-characteristics. So at lower level we construct a pair-wise comparison table for each of the attributes, reflecting expert knowledge on how much does a metric influence each attribute. Then, by applying normalization and the Eigenvalues extraction to each matrix, we obtain the each metric's weight for calculating the score for every attribute. This is repeated at the intermediate level of the hierarchy. Finally, at the top level, a pair-wise comparison table is also constructed reflecting expert knowledge of how much does each sub-characteristic influence its respective quality characteristic. The weights are also calculated by normalization and the Eigenvalues extraction.

The values for each entity for the ISO/IEC-9126 quality characteristics are calculated using the following utility function $U(C_i)$:

| | |
|---|---|
| $U(C_i) = v(sc_1)*w(sc_{1i}) + v(sc_2)*w(sc_{2i}) + ... + v(sc_n)*w(sc_{ni})$ | Equation (1), where |
| $v(sc_i) = v(d_1)*w(d_{1i}) + v(d_2)*w(d_{2i}) + ... + v(d_n)*w(d_{ni})$ | Equation (2), and |
| $v(d_i) = v(m_1)*w(m_{1i}) + v(m_2)*w(m_{2i}) + ... + v(m_n)*w(m_{ni})$ | Equation (3) |

| |
|---|
| $U(C_i)$ = Utility Function of ISO/IEC-9126 characteristic $I$ |
| $v(sc_i)$ = Value of Sub-characteristic $j$ |
| $w(sc_{ji})$ = Weight of Sub-characteristic $j$ for ISO/IEC-9126 Characteristic $i$ |
| $v(d_i)$ = Value of Source Code Attribute $d_i$ |
| $w(d_{ji})$ = Weight of Source Code Attribute $d_{ji}$ for Sub -Characteristic $i$ |
| $v(m_i)$ = Value of Metric $m_i$ |
| $w(m_{ji})$ = Weight of Metric $m_j$ for  Attribute $i$ |

## 4. EVALUATION

The proposed methodology was evaluated based on the following criteria:

- It should be flexible; that means to be suitable for evaluating systems with different functionality, developed using either the structured or the object oriented paradigm.

- It should be valid; that means to reflect the views and intuitions of domain experts concerning system level quality.

Based on the above criteria, the assessment of the methodology involved the following case studies:

- The analysis of two open source application servers, the Apache Geronimo and the JBoss AS, both developed in Java. The aim of this case study was to evaluate the ability of the proposed methodology to reflect the system trends in terms of ISO/IEC-9126 quality characteristics. The release notes for each version of these systems were used in order to validate the outcome of our methodology.

The maintainability evaluation of four software libraries, three open-source and one proprietary, used for developing telecommunication products based on the Session Initialisation Protocol (SIP). These libraries are developed either in C, C++ or a mix of both. The aim of this case study was to help the software engineers of a systems on-a-chip supplier, to select among these libraries the one that would be most suitable with regards to reuse, for the development of a soft-phone. The outcome of this evaluation was validated by the software engineers based on their perception and intuition of the maintainability levels of the four libraries.

## 4.1. Case Study 1: ISO/IEC-9126 Quality Characteristics Trend Analysis

As the proposed methodology employs source code attributes for assessing ISO/IEC-9126 characteristics, these should indicate trends that are required by systems that exhibit a certain level of quality. These trends are related with the laws of software evolution [21]. Firstly, functionality and efficiency are expected to increase from one version to the next as new features are added and capabilities are extended in order to incorporate additional requirements. In this way a software system remains adaptable to continuous changes and satisfactory in use. This is compatible with the laws of continuing change and growth of software systems [21].

However, maintainability is expected to decrease initially as new features are added and capabilities are extended by adding new classes and methods, thus making source code more complex. This is normal according to the law of the increasing complexity, which states that, as software systems evolve their complexity increase unless work is done to maintain or reduce it [21]. On the other hand when most of the required capabilities of a software system have been incorporated, maintenance effort focuses on reducing code and design complexity.

We evaluated Apache Geronimo, version 1.0 [22] and JBoss AS, version 4.0 [23]. Table 2 presents the size of the two software systems, measured in classes and in lines of code.

Table 2. Size of Apache Geronimo – JBoss AS

| Apache Geronimo Ver. | # Classes | KLOC | Release Date | JBoss AS Ver. | # Classes | KLOC | Release Date |
|---|---|---|---|---|---|---|---|
| 1.0 | 1650 | 166 | 05/01/2006 | 4.0.0 | 5907 | 283 | 20/09/2005 |
| 1.1 | 1642 | 161 | 26/06/2006 | 4.0.2 | 5885 | 288 | 02/05/2005 |
| 1.1.1 | 1657 | 164 | 18/09/2006 | 4.0.4 | 6887 | 325 | 15/05/2006 |
| | | | | 4.0.5 | 6517 | 296 | 18/10/2006 |

Table 3. Case Study 1 Selected Metrics

| Metric | Definition | Purpose |
|---|---|---|
| WMC | Sum of the cyclomatic complexity [24] of each member method of the class [20] | It estimates a class' maintainability and reusability by measuring its complexity |
| LCOM | No. of a class' member methods that access the same member data [13] | It measures if a class' methods work together to achieve a single purpose |
| CBO | No. of invocations of other classes' member methods or instance variables | It represents the no. of classes coupled to a given class |
| DIT | Max distance from top | It measures a class' inheritance levels from the object hierarchy top |
| NOC | No. of a class' immediate descendants | It measures the no. of a class' immediate descendants |
| NOM | No. Of Messages | It measures the services provided by a class |
| DSC | Total no. of classes [8] | It measures the system scale |
| DAM | Ratio of the no. of private and protected member data to the total no. of member data declared in a class | It reflects how well the property of encapsulation is applied to a class [8] |
| MOA | No. Of member data declarations whose types are user defined classes | It measures the extent of the part-whole relationship realized by using attributes [8] |
| NOP | No. Of the member methods that exhibit Polymorphic behaviour | It measures the overridden (or virtual) methods of an OO system |

In order to depict the actual trends of the ISO/IEC-9126's characteristics of versions 1.0 and 4.0 of Apache Geronimo and JBoss AS respectively, we normalised the derived characteristics' values according to the first version of each system. More specifically, the values of each characteristic were divided by the characteristics' values in the first version.

### 4.1.1. Selected Metrics

Table 3 outlines the selected metrics and their purpose when evaluating a software system's quality according to ISO/IEC-9126. It should be noted that all the proposed metrics apply to a system's classes except from Design Size in Classes (DSC), which applies to the system level.

### 4.1.2. Assigning Metrics to Source Code Attributes

According to ISO/IEC-9126 a source code attribute may contribute to one or more characteristics, and a characteristic consists of specific sub-characteristics [2]. One of the steps during the development of our quality evaluation methodology is to define an appropriate set of metrics for the evaluation of the source code attributes. Table 4 summarizes the assignment of the selected metrics presented in Table 3, to the attributes presented in Table 1.

Table 4. Case Study 1 Assigned Metrics

| Source Code Attribute | Metric |
|---|---|
| Volume | No. Of Statements |
| Complexity | Weighted Methods Per Class |
| Abstraction | No. Of Children |
| Encapsulation | Data Access Metric |
| Coupling | Coupling Between Objects |
| Cohesion | Lack of Cohesion in Methods |
| Messaging | No. of Messages |
| Polymorphism | No. of Polymorphic Methods |
| Composition | Measure of Aggregation |
| Inheritance | Depth Of Inheritance Tree |

### 4.1.3. Weights Assignment

The last step of the methodology for this case study was the assignment of weights to source code attributes in order to reflect their importance when evaluating ISO/IEC-9126 quality characteristics. Table 5 shows the weight of each attribute.

Table 5. Source Code Attributes Assigned Weights

| Attribute | Functionality | Efficiency | Maintainability | Portability |
|---|---|---|---|---|
| Volume | 0.17 | 0.05 | -0.12 | -0.06 |
| Complexity | 0.10 | 0.07 | -0.12 | -0.10 |
| Abstraction | 0.05 | 0.13 | 0.12 | 0.16 |
| Encapsulation | 0.04 | 0.13 | 0.12 | 0.05 |
| Coupling | 0.07 | 0.08 | -0.12 | -0.16 |
| Cohesion | 0.10 | 0.06 | 0.12 | 0.06 |
| Messaging | 0.17 | 0.09 | 0.06 | 0.06 |
| Polymorphism | 0.17 | 0.12 | -0.12 | 0.16 |
| Composition | 0.07 | 0.13 | 0.04 | 0.04 |
| Inheritance | 0.07 | 0.13 | 0.04 | 0.16 |

In order to reflect the significance of the individual source code attributes to each quality characteristic Bansiya and Davis have proposed in their Q.M.O.O.D. model a set of proportionally weighted attributes [8]. A range from -1 to 1 was selected for the computed values of the source code attributes. Based on these proposed relationships we applied AHP to the levels of hierarchy presented in §3.1. In this case study we evaluated the source code attributes (complexity, coupling, etc) for each ISO/IEC-9126 characteristic. Hence, at first we constructed a pair wise comparison table for each one of the source code attributes reflecting how much each attribute influences each characteristic. Then, by applying normalization and Eigenvalues extraction to each matrix, we find the weight of each attribute by calculating a score for each characteristic. Finally, a pair wise comparison table is constructed reflecting expert knowledge of how much each source code attribute influences each ISO/IEC-9126 characteristic; the weights are calculated by normalization and Eigenvalues extraction.

The volume attribute is viewed to be closely related to functionality, maintainability and portability. It is difficult to migrate and maintain a large system, which in turn is expected to provide ample functionality. Complexity, on the other hand indicates a system's maintainability and portability [4]. The more complex a system is, the harder it is to comprehend it in order to maintain it or to install it in another platform.

Abstraction influences functionality, portability and efficiency, as it supports faster and safer programming [8]. It also favours maintainability as it allows ignoring irrelevant details. Encapsulation promotes maintainability and portability as it makes it easier to understand system structure. More specifically, it helps managing code complexity by forbidding looking at the complexity. Coupling is also related to maintainability and portability. The goal for a software system is to create classes with small, direct, visible and flexible relations to other classes, which is known as "loose coupling". Higher measurements of coupling negatively influence these two quality characteristics. Cohesion is related to functionality, maintainability and portability. A high value of cohesion is considered good.

Objects communicate by using messages and therefore this influences their functionality, efficiency and portability. The greater the number of methods that can be invoked from a class through messages, the more complex the class is. Polymorphism also influences the same characteristics, but it also makes it harder to comprehend a class's design [8]. On the other hand, composition is closely related to efficiency and portability as a carefully composed system leads to a more effective design and is much easier to understand and therefore to be migrated to another platform [8].

Inheritance favours functionality, efficiency and portability as it simplifies programming. Deeper inheritance trees would seem to promote greater method sharing than broader trees. On the other hand excessive use of hierarchy trees has the potential to adversely influence maintainability [8].

### 4.1.4. System Evaluation

Fig. 5 and 7, present the evolution of the ISO/IEC-9126 quality characteristics for Apache Geronimo and JBoss AS respectively. In these figures the measurement scale of X-axis is ordinal with reference to the version of each software system and on the Y-axis are the normalised values of ISO/IEC-9126 characteristics. Fig. 6 and 8 on the other hand depict the evolution of the attributes of those systems. On the Y-axis are the normalised values of systems' attributes, while the ordering on the X-axis is nominal (i.e. reference to model's source code attributes).

### 4.1.4.1. Apache Geronimo Evaluation

According to Fig. 5 the increase in the values of functionality, efficiency and portability is compatible with the requirements imposed in section 4.2. More specifically efficiency increases significantly in every version of Apache Geronimo, from 1.0 to 1.1.1. Interestingly, portability is significantly increasing in version 1.1 and is slightly increasing in version 1.1.1 in relation to the previous version (i.e. 1.1). According to the release notes of version 1.1, it introduces several structural changes in order to improve Geronimo's portability [25]. This can explain why portability has such an increase in this version. Functionality is slightly increased in the next version of Apache Geronimo.

However, maintainability does not conform to the hypothesis that it decreases as the software system takes additional functionality. Both 1.1 and 1.1.1 versions are slightly more maintainable than the first version. This implies that work has been done in order to control the evolution of the source code and keep it maintainable.

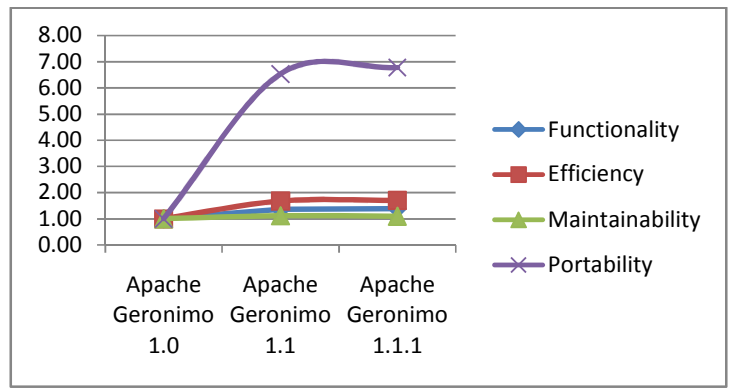

Figure 5. ISO/IEC-9126 Characteristics Evolution for Apache Geronimo

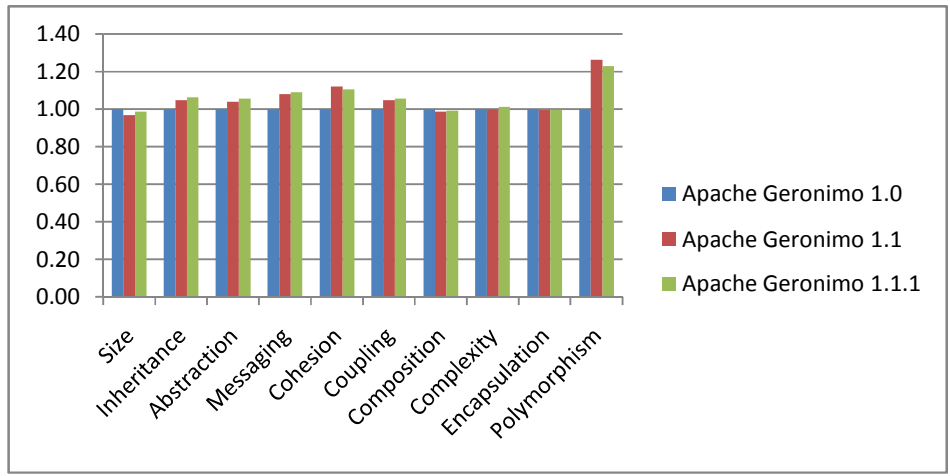

Figure 6. Apache Geronimo Attribute Evolution

Fig. 6 depicts that complexity, composition, encapsulation and volume have only slightly changed from one version to the other. The attributes with the most significant changes are polymorphism, cohesion and messaging and then coupling, inheritance and abstraction. The vast increase in the portability of the system was expected as it heavily depends on cohesion, polymorphism, inheritance and messaging, as described in Table 4. These factors see a notable increase from version to version, thus system portability improves. On the contrary, these factors contribute less towards the three other characteristics; this is why they only change marginally.

A useful observation can be made about the trends of the whole system at a higher level:

- While the volume of the system has not changed significantly; many attributes have been improved, like polymorphism, cohesion, messaging, and inheritance; thus system functionality has improved due to many changes rather than system extensions. The release notes of version 1.1 confirm these observations [25].

- Despite this improvement, the other attributes of the system have not improved any further, thus future system development would have to deal with this. Composition, complexity and encapsulation have to improve in order for the system as a whole to improve.

### 4.1.4.2. JBoss AS Evaluation

As seen in Fig. 7, the increase in the values of efficiency, portability and functionality is compatible with the requirements imposed in section 4.2. These quality characteristics increase from version 4.0.0 to 4.0.4 of JBoss AS. This fact can be explained by observing Fig. 8 depicting that, encapsulation, polymorphism, composition, inheritance, cohesion, coupling, volume and complexity vary significantly from version to version.

More specifically, functionality increases since attributes that mainly affect it, such as volume and especially polymorphism, also increase throughout. As far as efficiency is concerned, we see in Table 5 that it is equally affected by abstraction, composition, encapsulation, polymorphism and inheritance. Therefore, the big increase of the last three attributes mentioned, contributes to the improvement of the efficiency of JBoss from version 4.0.0 to 4.0.4.

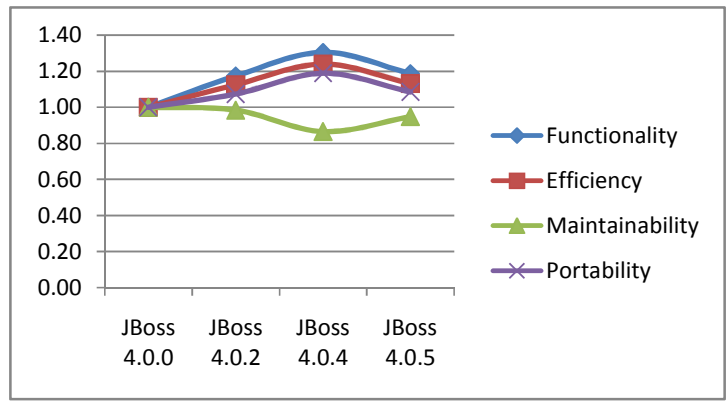

Figure 7. ISO/IEC-9126 Characteristics Evolution for JBoss AS

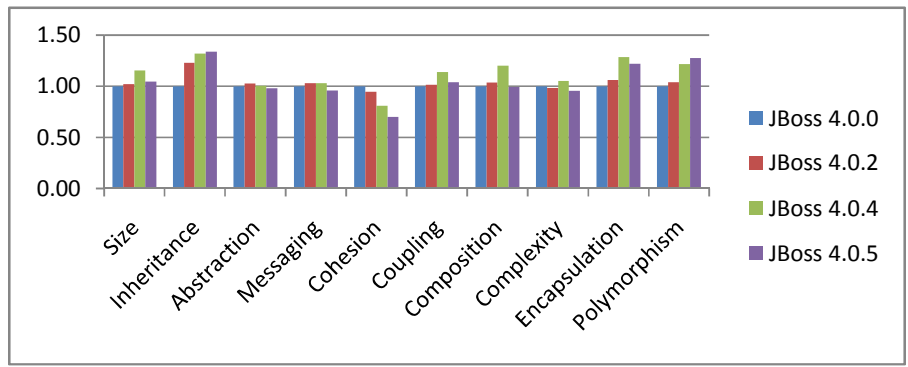

Figure 8. JBoss AS Source Code Attributes Evolution

In version 4.0.4 there is a significant change on the volume of JBoss and a lot of new features were added in order to satisfy change requests and to fix reported bugs from previous versions, according to this version's release notes [26]. Finally, portability mostly depends on abstraction, polymorphism and inheritance and it is negatively affected by coupling. Hence, since coupling slightly increases, while the increase of polymorphism and inheritance is much more significant, the portability of JBoss throughout versions improves, as expected.

Maintainability on the other hand, confirms the hypothesis that it decreases as the software system incorporates additional functionality; therefore it becomes more complex. More specifically it decreases in all versions except from 4.0.5 which is still less maintainable than 4.0.0. Observing Fig. 8 also leads to this conclusion, since maintainability is a linear combination of almost equally weighted factors, most of which affect negatively its value. In particular, increasing polymorphism and coupling whilst cohesion decreases leads to a less maintainable system.

If we compare the evaluation of Geronimo and JBoss we will see that all characteristics make similar progress except from one (portability in Geronimo and maintainability in JBoss). Also Geronimo seems to be more stable with regards to the variation of source code attributes, while the attributes of JBoss have greater perturbation.

### 4.2. Case Study 2: Maintainability Evaluation, Sip-Stack Selection

In this case study the proposed methodology was employed in order to perform a source code quality evaluation concerning the maintainability of four Session Initiation Protocol (SIP) software libraries. This assessment was performed on behalf of a company (for confidentiality reasons we will refer to this company as CompanyA) that is a supplier of system-on-a chip solutions for digital cordless telephony (voice and voice/data). Currently, CompanyA wants to develop a soft-phone for Voice over IP (VoIP) telephony based on SIP. SIP is a signalling protocol for Internet conferencing, telephony, events notification and instant messaging. SIP libraries and stacks that implement it, provide telecom software developers an interface to initiate and control SIP based sessions for their applications. In this study the following libraries were evaluated:

- A commercial system, which for confidentiality reasons will be called SystemA.
- WengoPhone, an open source SIP stack provided by Wengo [27].
- Libosip, an open source SIP stack.
- Sofia-Sip, an open source project started by Nokia

Table 6 presents the size of the four software systems measured in lines of code.

Table 6. Size of SIP Libraries and Stacks

| Name | Version | LOC | Programming Language |
|---|---|---|---|
| SystemA | 2.8.0 | 18,959 | C/C++ |
| WengoPhone | 2.1.2 | 90,645 | C/C++ |
| LibOSIP | 3.0.3 | 25,685 | C/C++ |
| Sofia-Sip | 1.12.7 | 67,448 | C/C++ |

The engineers of CompanyA selected the four systems as candidates for reuse in order to develop their own soft-phone. All these systems were providing similar functionalities, thus the sole criterion for the selection was their maintainability, particularly their ability to accept modifications. The engineers also needed to identify system parts that require modifications. As a result we employed our methodology in order to assess the changeability and analysability of the above systems.

Fig. 9 depicts the steps of the performed evaluation, which consisted of two iterations. At first the four systems were examined regarding their changeability as the main criterion. The outcome of this iteration was the selection of two of them for the final round. Here, the two more changeable libraries were assessed in terms of their analysability.

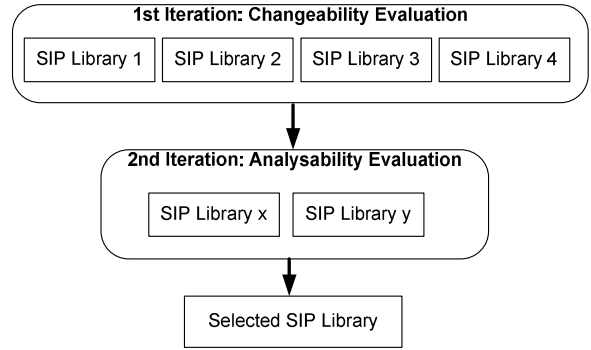

Figure 9. SIP Libraries Maintainability Evaluation Steps

### 4.2.1. Steps Followed

The aim of this case study was to assess the SIP libraries' capability in terms of their:

- Analysabillty, which is defined as the systems' capability to allow identification for parts which should be modified [2].

- Changeability, which is systems' capability to enable a specified modification (in our case changes to source code), to be implemented [2].

Those two maintainability sub-characteristics are influenced by certain source code attributes, which in turn are mapped into specific source code metrics. These relationships are depicted in Fig. 10.

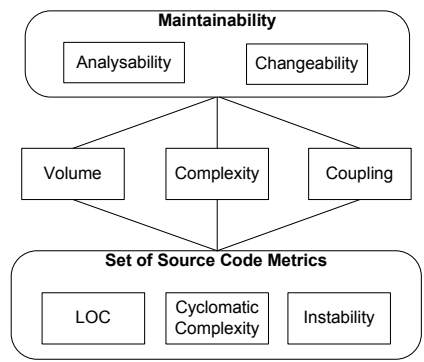

Figure 10. Case Study 2 Hierarchy

The influence of the attributes on changeability and analysability is as follows:

- Volume: It mainly influences a system's analysability [11], as a shorter piece of code is more readable and therefore easier to determine whether it requires a modification or not. Moreover, a large function or method body often indicates that the code performs many different and possibly unrelated tasks. In that case a system's maintainability is deteriorated [20].

- Complexity: The source code complexity is mainly influencing a system's changeability [11] and then its analysability. In a complex module it is not easy to identify the elements to change or to calculate the extension of the required changes. On the other hand a complex module is making difficult the task to locate either the causes of a failure or the software parts that are to be modified.

- Coupling: It mainly influences analysability and then changeability. A highly coupled module is difficult to understand as its dependencies disrupt reading its source code in order to locate dependency's target. Moreover coupling makes difficult for a module to identify to what extend the implemented changes affect the rest of the system.

Table 7, outlines the selected metrics and their purpose when evaluating software systems' maintainability according to ISO/IEC-9126. All the proposed metrics are applied on the methods of the four systems. Table 8 summarizes the assignment of the metrics presented in Table 7, to the source code attributes selected for applying our methodology.

The final step of the methodology is the assignment of weights to source code attributes in order to reflect their importance on evaluating the characteristics of ISO/IEC-9126. Table 9 shows the weight of each source code attribute on evaluating the two ISO/IEC-9126's maintainability sub-characteristics, changeability and analysability. Those weights were elicited based on the feedback and expertise of the software engineers of CompanyA.

Table 7. Case Study 2 Selected Metrics

| Metric | Definition | Purpose |
|--------|------------|---------|
| Lines Of Code (LOC) | LOC= no. of a function's / method's lines of code | It indicates the volume of a system |
| Cyclomatic Complexity (CC) | CC= e-n+2p<br>e: no. of sides, n: no. of edges<br>p: no. of adjacent components | It measures the level of complexity of the software design and coding structure [24] |
| Instability (I) | I= FO / (FO + FI)<br>FI=Fan In<br>FO=Fan Out | A measure of a function's total efferent (outgoing) couplings. It ranges from 0, (very stable) to 1 (very unstable) |

Table 8. Case Study 2 Assigned Metrics

| Source Code Attribute | Metric |
|-----------------------|--------|
| Volume | LOC |
| Complexity | Cyclomatic Complexity |
| Coupling | Instability |

Table 9. Case Study 2 Source Code Attributes Assigned Weights

| Source Code Attribute | Changeability | Analysability |
|-----------------------|---------------|---------------|
| Volume | 0.05 | 0.50 |
| Complexity | 0.80 | 0.25 |
| Coupling | 0.15 | 0.25 |

### 4.2.2. Systems Evaluation

The four SIP libraries were of different size as depicted in Table 6. For this reason, summation of the methods' measurement values (for complexity and instability) would not be helpful as it was strongly correlated to their volume [11]. Therefore in order to compare the four systems we computed relative volumes for each one by performing an aggregation at the system level. More specifically, we used two categorisations, one for complexity and the other for instability, presented in Tables 10 and 11 respectively. Then, we aggregated the complexity and instability per method, by counting for each level what percentage of lines of code falls within methods that belong to that level. For instance if, in a 10K LOC system, the high complexity risk units together amount to 500 LOC, then the computed aggregate number for that risk category is 5%. The same applied for coupling risk levels too.

Table 10. Complexity Categorisation

| Cyclomatic Complexity | Complexity Risk Level |
|---|---|
| 0-5 | Very low |
| 6-10 | Low |
| 11-20 | Moderate |
| 21-50 | High |
| >50 | Very High |

Table 11. Instability Categorisation

| Instability | Coupling Risk Level |
|---|---|
| 0.00-0.20 | Very low |
| 0.20-0.40 | Low |
| 0.40-0.60 | Moderate |
| 0.60-0.80 | High |
| 0.80-1.00 | Very High |

The first iteration of this case study was the evaluation of systems' changeability. Fig. 11 shows that Libosip and Sofia-Sip are the easier systems to change. Those two systems are less complex and more stable than SystemA and Wengo. This is depicted in Fig. 12 and 13.

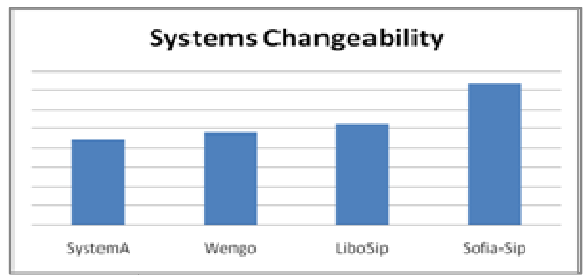

Figure 11: System Changeability Results

The second iteration was the evaluation of Libosip and Sofia-Sip in terms of their analysability. As Fig. 14 shows, Libo-Sip was the one easier to analyse and therefore the chosen library for reuse in order to build CompanyA's soft-phone. The main reason for this choice was the volume of Libo-Sip which is smaller than Sofia-Sip's. The software engineers of CompanyA were validated the final results. Their final comments were that the presented methodology was successfully presented in a quantitative way what was their intuition concerning the levels of the systems' maintainability.

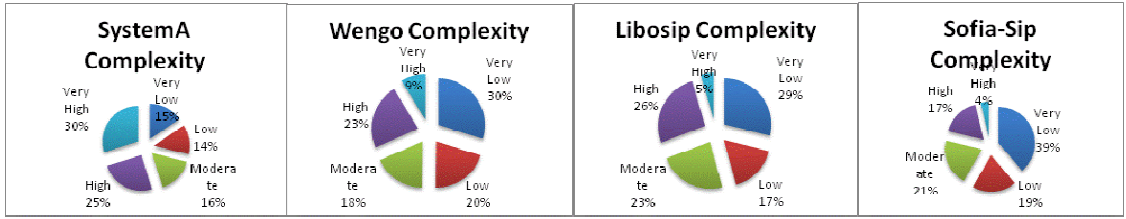

Figure 12. System Complexity

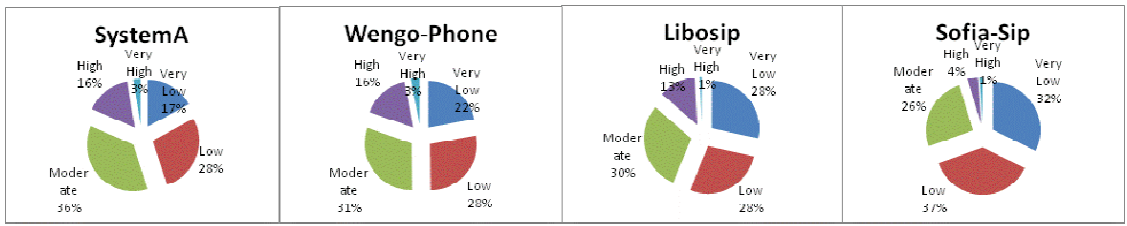

Figure 13. System Instability

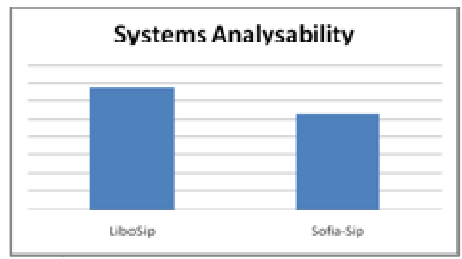

Figure 14. System Analysability Results

## 5. CONCLUSIONS

The aim of this work was to develop of a methodology for source code evaluation. This methodology uses as a frame of reference the ISO/IEC-9126 international standard for software quality that has been agreed upon by a majority of the international community. The proposed methodology enhances the hierarchy of this standard by introducing a layer which reflects the attributes of a software system's source code, such as volume, size, complexity, cohesion, coupling etc., similarly to the work of [11]. These attributes directly influence the characteristics and sub-characteristics of ISO/IEC-9126. The difference is that our work does not focus only on maintainability but also on portability, functionality and efficiency. In this way it is similar to the work of [12], which however does not use this intermediate source code properties layer.

Another characteristic of the proposed methodology is that it employs the AHP for the assignment of relative weights to metrics and source code attributes in order to reflect their importance when evaluating ISO/IEC-9126's characteristics and sub-characteristics. By using this method of multicriteria decision making, and using certain elements of subjectiveness, such as expert opinion, our methodology can make recommendations that approach the model's entities and relationships with more accuracy.

Our methodology consists of four steps. The first step involves the selection of specific ISO/IEC 9126 quality standard characteristics that depend solely on source code's internal quality and static behaviour. For this reason we included only four characteristics (functionality, efficiency, maintainability, portability) and their respective sub-characteristics.

The second step involves identification of source code attributes that directly influence the characteristics and sub-characteristics selected during the first step. These attributes should encompass all the aspects of a software system independently of the paradigm (structured or object oriented) used for its development.

The next step of our methodology entails the selection of metrics suitable for the evaluation of structured or object-oriented systems. The final step requires weight assignment to source code attributes, reflecting their importance when evaluating ISO/IEC-9126 characteristics and sub-characteristics.

The methodology was evaluated in terms of flexibility, in other words its suitability for evaluating systems with varying functionality, architecture and development paradigm (structured or object oriented). We also evaluated its accuracy in terms of its ability to reflect domain expert opinion and intuition concerning system level quality. For this reason two case studies were conducted.

In the first case study, two versions of two open source Application Servers, the Apache Geronimo and JBoss AS, were evaluated. The outcome of this case study was that the proposed methodology can detect how the quality characteristics of a software system evolve. More specifically, for the Apache Geronimo, it detected that functionality was improved not by extending the system, but by performing changes on the existing source code. The release notes of the version 1.1 confirmed this observation. In addition, for JBoss AS it was observed that functionality and efficiency values were maximised, while maintainability reached its minimum in version 4.0.4. According to the release notes of this version there was a significant change on the volume of JBoss and a lot of new features were added in order to satisfy change requests and to fix reported bugs from previous versions [26].

The second case study involved the maintainability evaluation and selection of one of four C/C++ SIP libraries, three open source and one proprietary, in order to be used for the development of a soft-phone by a semi-conductors vendor. The outcome of this case study was the selection of one of these libraries, Libo-Sip. This library proved to be the most maintainable in terms of its ability to accept modifications and to identify parts of it that need to be modified. The vendor's software engineers validated the derived results. Their final comments were that the presented methodology was successfully presented in a quantitative way what was their intuition concerning the levels of the systems' maintainability.

The results of these case studies indicate that the proposed methodology has considerable merit concerning the evaluation of source code quality in terms of the ISO/IEC-9126 standard. However, there are various alternatives that can be considered for the enhancement of the proposed methodology. At first more experiments are required in order to further calibrate the weights that reflect the importance of metrics and source code properties to the ISO/IEC-9126 quality characteristics and sub-characteristics. These weights are subjective, in the sense they are based on the intuition and opinion of system experts. Thus, different people, with different needs and backgrounds can give different weights. Furthermore, this subjectivity is an inherent characteristic of AHP. Another direction then for future work can be research on how other multi-criteria decision aid techniques or refinements of AHP can be used for the task of weights assignment.

## ACKNOWLEDGEMENTS

This work has been partially supported by the Greek General Secretariat for Research and Technology and Dynacomp S.A. within the program "P.E.P. of Western Greece Act 3.4".

**Authors**

Short Biography

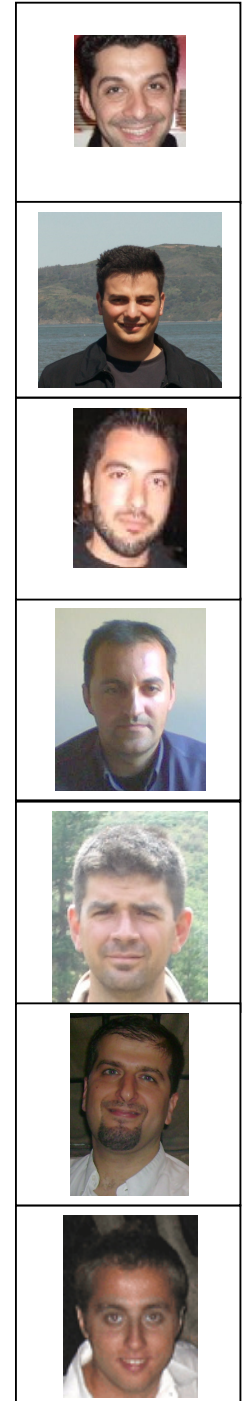

Yiannis Kanellopoulos is a senior consultant at SIG, based in the Netherlands. He is responsible for carrying out software quality and risk assessments for corporate and public clients. He holds an MSc in Information Systems Engineering and a PhD from University of Manchester, School of Computer Science. His PhD Thesis was related to the application of data mining techniques for supporting software systems maintenance. He has ample experience in managing software implementation projects.

Panos Antonellis is a Computer Engineer and a PhD student at the Dept. of Computer Engineering and Informatics, School of Engineering, Univ. of Patras.

Dimitris Antoniou is a Computer Engineer and Researcher in the Dept. of Computer Engineering and Informatics at the University of Patras. He has obtained his diploma from the Department in 2004 and his MSc in 2006. Since 2006, he has been a Ph.D. student at the same Dept. His research interests focus on Data Structures, Information Retrieval, String algorithmics, and bioinformatics, Software Quality Assessment, Web Technologies and GIS. He has scientific work published in int'l journals and conferences.

Christos Makris was born in Greece, in 1971. He graduated from the Dept. of Computer Engineering and Informatics, University of Patras, in December 1993. He received his Ph.D. degree from the Dept. of Computer Engineering and Informatics, in 1997. He is now an Assistant Professor in the same Department. His research interests include Data Structures, Web Algorithmics, Computational Geometry, Data Bases and Information Retrieval. He has published over 60 papers in scientific journals and refereed conferences.

Evangelos Theodoridis was born in Greece, in 1978. He graduated from the Dept. of Computer Engineering and Informatics, School of Engineering, University of Patras, in December 2002. He received his MSc. and a PhD degree from the same Dept. His research interests include Data Structures, Web Algorithmics, Data Bases and Information Retrieval and Bioinformatics.

Christos Tjortjis is an adjunct Senior Lecturer at the University of Ioannina, Dept. of Computer Science and the University of W. Macedonia, Dept. of Engineering Informatics and Telecoms' and an hon. Lecturer at the University of Manchester, School of Computer Science. He holds an MEng in Computer Engineering and Informatics from Patras, a BA in Law from Thrace, an MPhil in Computation from UMIST and a PhD in Informatics from Manchester. His research interests are in data mining, software comprehension and maintenance.

Nikos Tsirakis is a Computer Engineer and Researcher at the Dept. of Computer Engineering and Informatics at the University of Patras. He obtained his B.Eng. from the Dept. in 2004 and an MSc in 2006. Since 2006 he is a PhD student at the same Dept. His research focuses on String algorithmics and data structures, Hypertext modelling and searching, Software Quality Assessment, Web Technologies and GIS. He has scientific work published in int'l journals and conferences, and he co-authored books & encyclopaedia chapters.